\documentclass[onecolumn,prd,aps,12pt]{revtex4}
\usepackage{amsmath}
\usepackage{amsfonts}
\usepackage{amssymb}
\usepackage{graphicx}

\begin{document}

\title{Casimir Effect at finite temperature for the CPT-even extension of QED%
}
\author{ L. M. Silva$^{a}$, H. Belich$^{b}$, J. A. Helay\"{e}l-Neto$^{c}$.}
\affiliation{$^{a}${\small {\ Departamento de Ci\^{e}ncias Exatas e da Terra,
Universidade do Estado da Bahia, Campus II, 48040210 , Alagoinhas, BA,
Brazil.}}}
\affiliation{$^{b}${\small {Departamento de F\'{\i}sica e Qu\'{\i}mica, Universidade
Federal do Esp\'{\i}rito Santo, Vit\'{o}ria, ES, 29060-900, Brazil.}}}
\affiliation{{\small \ }$^{c}${\small {Centro Brasileiro de Pesquisas F\'{\i}sicas, Rua
Xavier Sigaud 150, Rio de Janeiro, RJ, 22290-180, Brazil.}}}
\email{belichjr@gmail.com, lourivalmsfilho@uol.com.br, helayel@cbpf.br. }
\date{\today}

\begin{abstract}
By the thermofield dynamics (TFD) formalism we obtain the energy-momentum
tensor for the Electromagnetism with Lorentz Breaking Even term of the
Standard Model Extended (SME) Sector in a topology $S^{1}\times S^{1}\times
R^{2}$. We carry out the compactification by a generalized TFD-Bogoliubov
transformation that is used to define a renormalized energy-momentum tensor,
and the Casimir energy and pressure at finite temperature are then derived.
A comparative analysis with the electromagnetic case is developed, and we
remark the influence of the background in the traditional Casimir effect.
\end{abstract}

\maketitle

\section{Introduction}

The Standard Model of particle physics (SM) has gotten success in its final
test: the discovery of the Higgs boson at the LHC in 2013. Despite the
tremendous success of this model, there is a fundamental question to be
answered, which is the justification for the Higgs mass. There is another
problem related to the mass of the Higgs boson: as the energy grows beyond
the SM energy scale, the radiative corrections makes the mass of the Higgs
boson to diverge (the problem of the Hierarchy). Also there is a lack of any
explanation for Dark Matter, and the unbalance between matter-antimatter.
Also, recently, The Standard Model of particle physics (SM) has not gotten
success in explaining the origin of electron's electric dipole moment (EDM), 
$d_{e}$, and its experimental upper bounds \cite{revmod}. Theories beyond
the SM predict a small, but potentially measurable EDM ($d_{e}\leq
10^{-29}\,e\cdot cm$) \cite{science}, which presents an asymmetric charge
distribution along the spin axis. Therefore, with this experimental result,
it is necessary to investigate the physics beyond the Standard Model.
Despite the great success of SM to give an overview of the microscopic
processes through a field theory that unifies the weak and electromagnetic
interaction SM presents nowadays some limitations.

For view of these limitations of the SM, one is greatly motivated to propose
models that\ can give us hints about a more fundamental theory. In 1989, in
a string theory environment, Kosteleck\'{y} and Samuel \cite{sam} realized
an interesting possibility in order to establish the spontaneous violation
of symmetry through non-scalar fields (vacuum of fields that have a tensor
nature) based on a string field theory environment. A consistent description
of fluctuations around this new vacuum is obtained if the components of the
background field are constant, and by the fact that this new minimum be a
background not scalar, the Lorentz symmetry is spontaneously broken \cite%
{ens}. This possibility of extending the Standard Model was made for fields
that belong to a more fundamental theory in which, in turn, can be
spontaneously violated based on a specific criterion. It is desirable that
any extension of the model could keep the gauge invariance, the conservation
of energy and momentum, and the covariance under observer rotations and
boosts. This proposal was known as Standard Model Extension (SME) \cite{col}
.

It is well-known that the presence of terms that violate the Lorentz
symmetry imposes at least one privileged direction in the spacetime.
Nowadays, studies in relativistic quantum effects \cite{knut1} that stem
from a non minimal coupling with Lorentz symmetry breaking \cite{knut2} has
opened the possibility of investigating new implications in quantum
mechanics that this violator background can promote.

An interesting question is investigate the influence of privileged
directions from Spontaneous Lorentz Violations coupled with the zero modes
of the electromagnetic field, i. e., in which way the vacuum fluctuation can
be affected?

The Casimir effect is one of the most remarkable manifestations of vacuum
fluctuations, and for the electromagnetic field, it consists in the
attraction between two metallic plates, parallel each other, embedded\ into
the vacuum \cite{Casimir48}. The attraction is due to a fluctuation of the
fundamental energy of the field caused by the presence of planes, which
select the electromagnetic vacuum modes by boundary conditions \cite%
{Milton,Miloni,Plunien,Mostepanenko}. In general, the Casimir effect is then
a modification in the vacuum energy of a given quantum field due to the
imposition of boundary conditions or topological effects on this field. The
measurements of this effect in great accuracy in the last decade has gained
attention of the theoretical and experimental community \cite{Lamo, Mohi}.
One practical implication of these achievements is the development of
nanodispositives. \cite{kempf,bordag,mtdo1}.{\Large \ }

In this paper we focus on the possibility of electromagnetism with breaking
of Lorentz symmetry in such way that the CPT symmetry is preserved. The even
sector of the SM\ with a vector decomposition is taken into account \cite%
{curv3, curv2}, and we analyze the influence on such breaking of symmetry in
the Casimir effect with thermal field treatment.

Considering that a thermal field theory is a quantum field compactified in a
topology $S^{1}\times R^{3}$, a result of the KMS (Kubo, Martin, Schwinger)
condition, this apparatus has been used to describe field theories in
toroidal topologies \cite{birrel1,birrel2,3comp1,3comp2,3comp3,3comp4}. In
terms of TFD, the Bogoliubov transformation has then been generalized to
describe thermal and space-compactification effects with real (not
imaginary) time. Here we consider a Bogoliubov transformation to take into
account the Lorentz spontaneous violation in a topology $S^{1}\times
S^{1}\times R^{2}$. Such a mechanism is quite suitable to treat, in
particular, the Casimir effect. This is a consequence of the nature of the
propagator that is written in two parts: one describes the flat (Minkowsky)
space-time contribution, whilst the other addresses to the thermal and the
topological effect. In such a case, a renormalized energy-momentum tensor is
introduced in a consistent and simple way \cite{kha1}. For the Casimir
effect, it is convenient to work with the real-time canonical formalism.

We have organized this paper in the following way: in Section II, some
aspects of TFD are presented to describe a field in a topology $S^{1}\times
S^{1}\times R^{3}$. In Section III, the energy-momentum tensor of our model
is derived. In Section IV, the topology $S^{1}\times S^{1}\times R^{3}$ is
considered; and Section V, the Casimir effect for our model is studied.
Concluding remarks are presented in Section VI.

\section{Thermofield dynamics and topology $S^{1}\times S^{1}\times R^{3}$}

In accordance with the ref. \cite{Leo} we present elements of thermofield
dynamics(TFD), emphasizing aspects to be used in the calculation of the
Casimir effect for our model. In short, TFD is introduced by two basic
ingredients \cite{kha1}. Considering a von-Neumann algebra of operator in
Hilbert space, there is a doubling, corresponding to the commutants
introduced by a modular conjugation. This corresponds to a doubling of the
original Fock space of the system leading to the expanded space $\mathcal{H}%
_{T}=\mathcal{H}\otimes \widetilde{\mathcal{H}}$. This doubling is defined
by a mapping $\widetilde{}:\mathcal{H}\rightarrow \widetilde{\mathcal{H}}$,
associating each operator $a$ acting on $\mathcal{H}$ with two operators in $%
\mathcal{H}_{T}$, $A\ $ and $\tilde{A}$, which are connected by the modular
conjugation in a c$^{\ast }$-algebra, also called tilde conjugation rules 
\cite{kha2,kha3}: 
\begin{align*}
(A_{i}A_{j})\widetilde{}& =\widetilde{A}_{i}\widetilde{A}_{j}, \\
(cA_{i}+A_{j})\widetilde{}& =c^{\ast }\widetilde{A}_{i}+\widetilde{A}_{j}, \\
(A_{i}^{\dagger })\widetilde{}& =(\widetilde{A}_{i})^{\dagger }\text{ },(%
\widetilde{A}_{i})^{\widetilde{}}=-\xi A_{i},
\end{align*}%
with $\xi =-1$ for bosons and $\xi =+1$ for fermions. The physical variables
are described by nontilde operators. The tilde variables, defined in the
commutant of the von Neumann algebra, are associated with generators of the
modular group given by $\widehat{A}=A-\widetilde{A}$. With this elements,
reducible representations of Lie-groups can be studied, in particular,
kinematical symmetries as the Lorentz group. This gives rise to
Liouville-von-Neumann-like equations of motion. The other basic ingredient
of TFD is a Bogoliubov transformation, $B(\alpha )$, introducing a rotation
in the tilde and non-tilde variables, such that thermal effects emerge from
a condensate state. The rotation parameter $\alpha $ is associated with
temperature, and this procedure is equivalent to the usual statistical
thermal average. \ In the standard doublet notation \cite{3ume2}, we write 
\begin{equation}
(A^{r}(\alpha ))=\left( 
\begin{array}{c}
A(\alpha ) \\ 
\xi \widetilde{A}^{\dagger }(\alpha )%
\end{array}%
\right) =B(\alpha )\left( 
\begin{array}{c}
A \\ 
\,\xi \widetilde{A}^{\dagger }%
\end{array}%
\right) \,,
\end{equation}%
$(\,A^{r}(\alpha ))^{\dagger }=\left( A^{\dagger }(\alpha )\,,\,\widetilde{A}%
(\alpha )\right) \,$, with the Bogoliubov transformation given by 
\begin{equation}
B(\alpha )=\left( 
\begin{array}{cc}
u(\alpha ) & -v(\alpha ) \\ 
\xi v(\alpha ) & \,\,\,\,u(\alpha )%
\end{array}%
\right) ,  \label{BT}
\end{equation}%
where $\,u^{2}(\alpha )+\xi v^{2}(\alpha )=1$.

The parametrization of the Bogoliubov transformation in TFD is obtained by
setting $\alpha =\beta =T^{-1}$ and by requiring that the thermal average of
the number operator, $N(\alpha )=a^{\dagger }(\alpha )a(\alpha )$, i.e. $%
\langle N(\alpha )\rangle _{\alpha }=\langle 0,\tilde{0}|a^{\dagger }(\alpha
)a(\alpha )|0,\tilde{0}\rangle $, \ gives either the Bose or the Fermi
distribution, i.e 
\begin{equation}
N(\alpha )=\,\,\,v^{2}(\beta )=\left( e^{\beta \varepsilon }+\xi \right)
^{-1}\,.  \label{UV}
\end{equation}%
Here we have used, for the sake of simplicity of notation, $A\equiv a$ and $%
\widetilde{A}\equiv \widetilde{a}$, and 
\begin{equation*}
a=u(\alpha )a(\alpha )+v(\alpha )\,\widetilde{a}^{\dagger }(k,\alpha ),
\end{equation*}%
such that the other operators ($a^{\dag }(k),\widetilde{a}(k),\widetilde{a}%
^{\dagger }(k)$) can be obtained by applying the Hermitian or the tilde
conjugation rules. It is shown then that the thermal average, $\langle
N(\alpha )\rangle _{\alpha }$, can be written as $\langle N(\alpha )\rangle
_{\alpha }=\langle 0(\alpha )|a^{\dagger }a|0(\alpha )\rangle $, where $%
|0(\alpha )\rangle $ is given by $|0(\alpha )\rangle =U(\alpha )|0,%
\widetilde{0}\rangle $, with 
\begin{equation*}
U(\alpha )=\exp \{\theta (\alpha )[a^{\dag }\widetilde{a}^{\dag }-a%
\widetilde{a}]\}.
\end{equation*}

Let us consider the free Klein-Gordon field described by the Hamiltonian $%
\mathcal{H}=\frac{1}{2}\partial_{\alpha}\phi\partial^{\alpha}\phi-\frac{1}{2}%
m^{2}\phi^{2},$ in a Minkowski space specified by the diagonal metric with
signature $(+,---)$. The generalization of $U(\alpha)$ is then defined for
all modes, such that%
\begin{align*}
\phi(x;\alpha) & =U(\alpha)\phi(x)U^{-1}(\alpha), \\
\widetilde{\phi}(x;\alpha) & =U(\alpha)\widetilde{\phi}(x)U^{-1}(\alpha).
\end{align*}
Using a Bogoliubov transformation for each mode, we get \cite{kha1} 
\begin{equation*}
\phi(x;\alpha)=\int\frac{d^{3}k}{(2\pi)^{3}}\frac{1}{2k_{0}}\ [a(k;\alpha
)e^{-ikx}+a^{\dag}(k;\alpha)e^{ikx}]
\end{equation*}
and%
\begin{equation*}
\widetilde{\phi}(x;\alpha)=\int\frac{d^{3}k}{(2\pi)^{3}}\frac{1}{2k_{0}}\ [%
\widetilde{a}(k;\alpha)e^{ikx}+\widetilde{a}^{\dag}(k;\alpha)e^{-ikx}].
\end{equation*}

The $\alpha$-propagator is defined by 
\begin{align}
G(x-y,\alpha) & =-i\langle0,\widetilde{0}|\mathrm{T}[\phi(x;\alpha
)\phi(y;\alpha)]|0,\widetilde{0}\rangle  \notag \\
& =-i\langle0(\alpha)|\mathrm{T}[\phi(x)\phi(y)]|0(\alpha)\rangle,
\label{T3}
\end{align}
where T is the time-ordering operator. This leads to%
\begin{equation}
G_{0}(x-y,\alpha)=\int\frac{d^{4}k}{(2\pi)^{4}}e^{-ik(x-y)}\ G_{0}(k,\alpha),
\label{T2}
\end{equation}
where%
\begin{equation}
G_{0}(k;\alpha)=G_{0}(k)+\ v^{2}(k_{\alpha};\alpha)[G_{0}(k)-G_{0}^{\ast
}(k)],  \label{T1}
\end{equation}
with 
\begin{equation*}
G_{0}(k)=\frac{1}{k^{2}-m^{2}+i\varepsilon},
\end{equation*}
such that%
\begin{equation*}
G_{0}(k)-G_{0}^{\ast}(k)=2\pi i\delta(k^{2}-m^{2}).
\end{equation*}
Using $v^{2}(k_{\alpha};\alpha)=v^{2}(k^{0};\beta)$ as the boson
distribution, $n(k^{0};\beta)$, i.e. 
\begin{equation}
v^{2}(k^{0};\beta)=n(k^{0};\beta)=\frac{1}{(e^{\beta\omega_{k}}-1)}%
=\sum\limits_{l_{0}=1}^{\infty}e^{-\beta k^{0}l_{0}},  \label{june121}
\end{equation}
with $\omega_{k}=k_{0}$ and $\beta=1/T$, $T$ being the temperature, then we
have 
\begin{equation}
G(k,\beta)=G_{0}(k)+2\pi i~n(k^{0},\beta)\delta(k^{2}-m^{2}),  \label{mats71}
\end{equation}
with%
\begin{equation*}
G_{0}(x-y)=\int\frac{d^{4}k}{(2\pi)^{4}}e^{-ik(x-y)}\ G_{0}(k).
\end{equation*}
For the case $m=0$, we have 
\begin{equation}
G_{0}(x-y)=\frac{-i}{(2\pi)^{2}}\frac{1}{(x-x^{\prime})^{2}-i\varepsilon },
\label{june1}
\end{equation}
\ 

The Green function given in Eq. (\ref{T3}) is also written as 
\begin{align*}
G_{0}(x-y,\beta) & =\mathrm{Tr}[\rho(\beta)\mathrm{T}[\phi(x)\phi(y)]] \\
& =G_{0}(x-y-i\beta n_{0},\beta),
\end{align*}
where $\rho(\beta)$ is the equilibrium density matrix for the
grand-canonical ensemble and $n_{0}=(1,0,0,0)$. This shows that $%
G_{0}(x-y,\beta)$ is a periodic function in the imaginary time, with period
of $\beta$; and the quantities $w_{n}=2\pi n/\beta$ are the Matsubara
frequencies. This periodicity is known as the KMS (Kubo, Martin, Schwinger)
boundary condition. From Eq. (\ref{T2}) we show that $G_{0}(x-y,\beta)$ is a
solution of the Klein-Gordon equation: with $\tau=it,$ such that $%
\square+m^{2}=-\partial_{\tau}^{2}-\nabla^{2}+m^{2}$, and 
\begin{equation}
(\square+m^{2})G_{0}(x,\beta)=-\delta(x).  \label{T7}
\end{equation}
Then $G_{0}(x-y,\beta)$ can also be written as a in a Fourier series, i.e. 
\begin{equation}
G_{0}(x-y,\beta)=\frac{-1}{i\beta}\sum_{n}\int d^{3}p\frac{e^{-ik_{n}\cdot x}%
}{k_{n}^{2}-m^{2}+i\varepsilon},  \label{8propmat1}
\end{equation}
where $k_{n}=(k_{n}^{0},\mathbf{k}).$ The propagator, given in Eq.~(\ref{T3}%
) and in Eq.~(\ref{8propmat1}), is solution of Eq.~(\ref{T7}) and fullfils
the same boundary condition of periodicity and Feyman contour. Then these
solutions are the same. A direct proof is provided by Dolan and Jackiw in
the case of temperature~\cite{DJ1}.

Due to the periodicity and the fact $G_{0}(x,\beta)$ and $G_{0}(x-y)$
satisfy Eq.~(\ref{T7}), the same local structure, then this finite
temperature theory results to be the $T=0$ theory compactified in a topology 
$\Gamma_{4}^{1}=S^{1}\times\mathbb{R}^{3}$, where the (imaginary) time is
compactified in $S^{1}$, with circumference $\beta$. The Bogoliubov
transformation introduces the imaginary compactification through a
condensate.

For an Euclidian theory, this procedure can be developed for space
compactification. In accordance with \cite{Leo} the Bogoliubov
transformation is given by 
\begin{equation}
v^{2}(k^{1},L_{1})=\sum\limits_{n=1}^{\infty }e^{-inL_{1}k^{1}}.
\label{june125}
\end{equation}

From this result, we compactify this theory in the imaginary time in order
to take into account the temperature effect. We consider now the topology $%
\Gamma _{4}^{2}=S^{1}\times S^{1}\times R^{2}$. The boson field is
compactified in two directions, i.e. $x^{0}$ and $x^{1}$. In the $x^{1}$%
-axis, the compactification is in a circle of circumference $L_{1}$ and in
the Euclidian $x^{0}$- axis, the compactification is in a circumference $%
\beta $, such that in both of the cases the Green function satisfies
periodic boundary conditions. In this case, the Bogoliubov transformation is
given by%
\begin{align}
v^{2}(k^{0},k^{1};\beta ,L_{1})& =v^{2}(k^{0};\beta )+v^{2}(k^{1};L_{1}) 
\notag \\
& +2v^{2}(k^{0};\beta )v^{2}(k^{1};L_{1}).  \label{june124}
\end{align}%
This corresponds to a generalization of the Dolan-Jackiw propagator,
describing a system of free bosons at finite temperature, with a
compactified space dimension~\cite{kha1,DJ1}. Observe the following
consistency relations 
\begin{align*}
v_{B}^{2}(k^{0};\beta )& =\lim_{L_{1}\rightarrow \infty
}v^{2}(k^{0},k^{1};\beta ,L_{1}), \\
v_{B}^{2}(k^{1};L_{1})& =\lim_{\beta \rightarrow \infty
}v^{2}(k^{0},k^{1};\beta ,L_{1}).
\end{align*}%
In the next sections we use these results to analyze the energy-momentum
tensor of our model.

\section{The Electromagnetism with Lorentz Breaking Even of SME Sector}

Now we begin this section analizing a modifief electromagnetism by a
CPT-even term of the SME. A way of investigating the effects of the
violation of the Lorentz symmetry is to consider parameters associated with
vector and tensor fields (background), such as $K_{\mu \nu \kappa \lambda }$%
. Based on Maxwell's electrodynamics, these background fields should be very
small because any effect associated with them is expected to be in an energy
scale that we have never accessed. Both vector and tensor fields are
considered to be background fields since they permeate the whole spacetime
and we have no access to its source, i.e., they are fixed vector and tensor
fields that select a privileged direction in the spacetime, and thus break
the isotropy. Thereby, based on Maxwell's electrodynamics, the Lagrangian of
the gauge sector of the Standard Model Extension is given by \cite{col}

We propose to carry out our analysis by starting off from the action 
\begin{equation}
\Sigma =\int d^{4}x\left\{ -\frac{1}{4}F_{\mu \nu }F^{\mu \nu }\right\}
+\Sigma _{k},  \label{1}
\end{equation}%
where $F^{\mu \nu }=\partial ^{\mu }A^{\nu }-\partial ^{\nu }A^{\mu }$ is
the eletromagnetic field strenght tensor.

\begin{equation}
\Sigma _{k}=-\frac{1}{4}\int d^{4}x\left( \kappa ^{\mu \nu \rho \sigma
}\,F_{\mu \nu }F_{\rho \sigma }\right)  \label{2}
\end{equation}%
where 
\begin{equation}
\mathcal{L}_{\kappa }=-\,\left( \frac{1}{4}\,\kappa ^{\mu \nu \rho \sigma
}\,F_{\mu \nu }F_{\rho \sigma }\right) \ ,
\end{equation}%
Then we deal with a Lorentz symmetry violating tensor $K_{\mu \nu \kappa
\lambda }$\ in such a way that the CPT symmetry is preserved. In recent
years, it has been shown in Ref. \cite{hugo} that a particular decomposition
of the tensor $K_{\mu \nu \kappa \lambda }$\ produces a modification on the
equations of motion of the electromagnetic waves due to the presence of
vacuum anisotropies, which gives rise to the modified Maxwell equations. As
a consequence, the anisotropy can be a source of the electric field and,
then, Gauss's law is modified. Besides, the Amp\`{e}re-Maxwell law is
modified by the presence of anisotropies and it has a particular interest in
the analysis of vortices solutions since it generates the dependence of the
vortex core size on the intensity of the anisotropy.

The \textquotedblleft tensor\textquotedblright\ $K_{\mu \nu \kappa \lambda }$
it's CPT-even, i. e., don't violates the CPT symmetry. Despite CPT violation
implies violation of Lorentz invariance , the reverse is not true. The
action above is Lorentz-violanting in the sense that the \textquotedblleft
tensor\textquotedblright\ $K_{\mu \nu \kappa \lambda }$ has a non-zero
vaccum expectation value. That \textquotedblleft tensor\textquotedblright\
presents the following symmetries:

\begin{equation}  \label{anz1}
K_{\mu \nu \kappa \lambda }=K_{\left[ \mu \nu \right] \left[ \kappa \lambda %
\right] },\; K_{\mu \nu \kappa \lambda }=K_{\kappa \lambda \mu \nu },\; K{_{{%
\mu \nu }}}^{\mu \nu }=0,
\end{equation}

as usually appears in the literature,\ we can reduce the degrees of freedom
take into account the ansatz :

\begin{equation}  \label{anz2}
K_{\mu \nu \kappa \lambda }=\frac{1}{2}\left( \eta _{\mu \kappa }\tilde{%
\kappa}_{\nu \lambda }-\eta _{\mu \lambda }\tilde{\kappa}_{\nu \kappa }+\eta
_{\nu \lambda }\tilde{\kappa}_{\mu \kappa }-\eta _{\nu \kappa }\tilde{\kappa}%
_{\mu \lambda }\right) ,
\end{equation}

\begin{equation}  \label{anz3}
\tilde{\kappa}_{\mu \nu }=\kappa \left( \xi _{\mu }\xi _{\nu }-\eta _{\mu
\nu }\xi ^{\alpha }\xi _{\alpha }/4 \right) ,
\end{equation}

\begin{equation}
\kappa =\frac{4}{3}\tilde{\kappa}^{\mu \nu }\xi _{\mu }\xi _{\nu },
\end{equation}

\bigskip Using the decomposition in the term Lorentz violating term we have,

\begin{equation}
\kappa ^{\mu \nu \rho \sigma }\,F_{\mu \nu }F_{\rho \sigma }=2\kappa \left(
g^{\mu \rho }\left( \xi ^{\nu }\xi ^{\sigma }-g^{\nu \sigma }\,\xi ^{e}\xi
_{e}/4\right) \right) F_{\mu \nu }F_{\rho \sigma }
\end{equation}

Then joint the Maxwell with the CPT-even violating term, we have,

\begin{equation}
\Sigma _{g}=\int d^{4}x\left\{ -\frac{1}{4}\left( 1-\frac{\kappa }{2}\xi
^{e}\xi _{e}\right) F_{\mu \nu }F^{\mu \nu }-\frac{\kappa }{2}\xi ^{\nu }\xi
^{\sigma }F\text{ }^{\rho }\text{ }_{\nu }F_{\rho \sigma }\right\}
\end{equation}

To calculate the Casimir effect for this model we need the expression of the
moment tensor-energy. With this expression we can evaluate the effect of
violating the background Casimir effect.

\section{The Energy-Momentum Tensor}

The complete expression of the energy-momentum tensor is given by:

\begin{equation}
\mathcal{T}_{\alpha \beta }=-\left( 1-\frac{\kappa }{2}\xi ^{2}\right) F%
\text{ }^{\rho }\text{ }_{\beta }F_{\rho \alpha }+g_{\alpha \beta }\frac{1}{4%
}\left( 1-\frac{\kappa }{2}\xi ^{2}\right) F_{\mu \nu }F^{\mu \nu }-\kappa
\xi ^{\nu }\xi ^{\sigma }F_{\alpha \nu }F_{\beta \sigma }+g_{\alpha \beta }%
\frac{\kappa }{2}\xi ^{\nu }\xi ^{\sigma }F\text{ }^{\rho }\text{ }_{\nu
}F_{\rho \sigma },
\end{equation}

and establishing the relationship $\mathcal{T}_{\alpha \beta }=\mathcal{T}%
_{\alpha \beta }^{a}+\mathcal{T}_{\alpha \beta }^{b},$where,

\begin{eqnarray}
\text{\ }\mathcal{T}_{\alpha \beta }^{a} &=&-\left( 1-\frac{\kappa }{2}\xi
^{2}\right) F\text{ }^{\rho }\text{ }_{\beta }F_{\rho \alpha }+g_{\alpha
\beta }\frac{1}{4}\left( 1-\frac{\kappa }{2}\xi ^{2}\right) F_{\mu \nu
}F^{\mu \nu };\text{ } \\
\mathcal{T}_{\alpha \beta }^{b} &=&-\kappa \xi ^{\nu }\xi ^{\sigma
}F_{\alpha \nu }F_{\beta \sigma }+g_{\alpha \beta }\frac{\kappa }{2}\xi
^{\nu }\xi ^{\sigma }F\text{ }^{\rho }\text{ }_{\nu }F_{\sigma \rho }.
\end{eqnarray}

Observing that the gauge potential satisfies the equation,%
\begin{equation}
\theta _{\mu \nu }A^{\nu }=\left( g_{\mu \nu }-\frac{\partial _{\mu
}\partial _{\nu }}{\Box }\right) A^{\nu }(x)=0\;,
\end{equation}%
with the conjugated four momomentum $\pi ^{\mu }=\frac{\partial \mathcal{L}}{%
\partial \left( \partial _{0}A_{\mu }\right) },$ i.e. $\pi ^{0}=0$ $\pi
^{i}=\partial ^{0}A^{i}-\partial ^{i}A^{0}$, and the comutation relation
obeys,%
\begin{equation}
\left[ A_{i}(\mathbf{x},t),\pi _{j}(\mathbf{x}{\acute{}},t)\right] =i\left[
\delta _{ij}-\frac{1}{\nabla ^{2}}\partial _{i}\partial _{j}\right] \delta
\left( \mathbf{x},\mathbf{x}{\acute{}}\right) ,\text{ }i,j,k=1,2,3.
\end{equation}

Now we begin to work with $\mathcal{T}_{\alpha \beta }^{b},$ 
\begin{equation}
\mathcal{T}_{\alpha \beta }^{b}=-\kappa \mathcal{F}^{\sigma }\text{ }%
_{\alpha \nu }\left( \mathbf{x}\right) \mathcal{F}^{\nu }\text{ }_{\beta
\sigma }\left( \mathbf{x}{\acute{}}\right) +g_{\alpha \beta }\frac{\kappa }{2%
}\mathcal{F}^{\sigma \rho }\text{ }_{\nu }\left( \mathbf{x}\right) \mathcal{F%
}^{\nu }\text{ }_{\rho \sigma }\left( \mathbf{x}{\acute{}}\right)
\end{equation}

\bigskip where%
\begin{equation}
\mathcal{F}^{\sigma }\text{ }_{\alpha \nu }\left( \mathbf{x}\right) =\xi
^{\sigma }F_{\alpha \nu };\text{ }\Sigma ^{\mu }\text{ }_{\nu }=\xi ^{\mu
}\partial _{\nu }.
\end{equation}

To compute de vacuum expectation value,

\begin{equation}
\mathcal{T}_{\alpha \beta }^{b}\left( x\right) =\lim_{x\rightarrow x{\acute{}%
}}\left[ -\kappa \mathcal{F}^{\sigma }\text{ }_{\alpha \nu }\left( \mathbf{x}%
\right) \mathcal{F}^{\nu }\text{ }_{\beta \sigma }\left( \mathbf{x}{\acute{}}%
\right) +g_{\alpha \beta }\frac{\kappa }{2}\mathcal{F}^{\sigma \rho }\text{ }%
_{\nu }\left( \mathbf{x}\right) \mathcal{F}^{\nu }\text{ }_{\rho \sigma
}\left( \mathbf{x}{\acute{}}\right) \right]
\end{equation}

\bigskip in another way,%
\begin{equation}
\mathcal{T}_{\alpha \beta }^{b}\left( x\right) =\lim_{x\rightarrow x{\acute{}%
}}\left[ -\kappa \mathfrak{F}^{\sigma }\text{ }_{\alpha \nu },\text{\textbf{%
\ }}^{\nu }\text{ }_{\beta \sigma }\left( \mathbf{x,x}{\acute{}}\right)
+g_{\alpha \beta }\frac{\kappa }{2}\mathfrak{F}^{\sigma \rho }\text{ }_{\nu }%
\mathbf{,}\text{ }^{\nu }\text{ }_{\rho \sigma }\left( \mathbf{x,x}{\acute{}}%
\right) \right]
\end{equation}

such that,

\begin{equation}
\mathfrak{F}^{\sigma }\text{ }_{\alpha \nu }\left( \mathbf{x,x}{\acute{}}%
\right) =T\left[ -\kappa \mathcal{F}^{\sigma }\text{ }_{\alpha \nu }\left( 
\mathbf{x}\right) \mathcal{F}^{\nu }\text{ }_{\beta \sigma }\left( \mathbf{x}%
{\acute{}}\right) \right]
\end{equation}

\bigskip where $T$ is the time ordering operator,

\begin{eqnarray}
\mathfrak{F}^{\sigma }\text{ }_{\alpha \nu }\left( \mathbf{x,x}{\acute{}}%
\right) &=&\left( \Sigma ^{\sigma }\text{ }_{\alpha }A_{\beta }\left( 
\mathbf{x}\right) -\Sigma ^{\sigma }\text{ }_{\beta }A_{\alpha }\left( 
\mathbf{x}\right) \right) \left( \Sigma {\acute{}}^{\nu }\text{ }_{\beta
}A_{\sigma }\left( \mathbf{x}{\acute{}}\right) -\Sigma {\acute{}}^{\nu }%
\text{ }_{\alpha }A_{\beta }\left( \mathbf{x}{\acute{}}\right) \right)
\theta \left( \mathbf{x}_{0}\mathbf{-x}_{0}{\acute{}}\right) +  \notag \\
&&+\left( \Sigma {\acute{}}^{\sigma }\text{ }_{\beta }A_{\alpha }\left( 
\mathbf{x}{\acute{}}\right) -\Sigma {\acute{}}^{\sigma }\text{ }_{\alpha
}A_{\beta }\left( \mathbf{x}{\acute{}}\right) \right) \left( \Sigma ^{\sigma
}\text{ }_{\alpha }A_{\beta }\left( \mathbf{x}\right) -\Sigma ^{\sigma }%
\text{ }_{\beta }A_{\alpha }\left( \mathbf{x}\right) \right) \theta \left( 
\mathbf{x}{\acute{}}\mathbf{-x}\right) ,
\end{eqnarray}

which can be written in terms of $\Gamma $ such that,

\begin{equation}
\Gamma ^{\sigma }\text{ }_{\alpha \nu },\text{\textbf{\ }}^{\nu }\text{ }%
_{\beta \sigma }\text{ }^{\lambda \psi }\left( \mathbf{x,x}{\acute{}}\right)
=\left( \delta _{\nu }^{\lambda }\Sigma ^{\sigma }\text{ }_{\alpha }-\delta
_{\alpha }^{\lambda }\Sigma ^{\sigma }\text{ }_{\nu }\right) \left( \delta
_{\sigma }^{\psi }\Sigma {\acute{}}^{\nu }\text{ }_{\beta }-\delta _{\beta
}^{\psi }\Sigma {\acute{}}^{\nu }\text{ }_{\sigma }\right) ,  \label{10}
\end{equation}

\bigskip with%
\begin{eqnarray}
\mathfrak{F}^{\sigma }\text{ }_{\alpha \nu },\text{\textbf{\ }}^{\nu }\text{ 
}_{\beta \sigma }\left( \mathbf{x,x}{\acute{}}\right) &=&\Gamma ^{\sigma }%
\text{ }_{\alpha \nu },\text{\textbf{\ }}^{\nu }\text{ }_{\beta \sigma }%
\text{ }^{\lambda \psi }\left( \mathbf{x,x}{\acute{}}\right) T\left[
A_{\lambda }\left( \mathbf{x}\right) ,A_{\psi }\left( \mathbf{x}{\acute{}}%
\right) \right] +  \notag \\
&&-n_{0\alpha }\delta \left( \mathbf{x}_{0}-\mathbf{x}_{0}{\acute{}}\right)
I_{\nu },\text{\textbf{\ }}^{\nu }\text{ }_{\beta \sigma }\left( \mathbf{x,x}%
{\acute{}}\right) +n_{0\nu }\delta \left( \mathbf{x}_{0}-\mathbf{x}_{0}{%
\acute{}}\right) I_{\alpha },\text{\textbf{\ }}^{\nu }\text{ }_{\beta \sigma
}\left( \mathbf{x,x}{\acute{}}\right) ,\text{ }  \label{tau}
\end{eqnarray}

unless of expression,%
\begin{equation}
I_{\alpha },\text{\textbf{\ }}^{\nu }\text{ }_{\beta \sigma }\left( \mathbf{%
x,x}{\acute{}}\right) =\left[ A_{\alpha }\left( \mathbf{x}\right) ,\mathcal{F%
}{\acute{}}^{\nu }\text{ }_{\beta \sigma }\left( \mathbf{x}{\acute{}}\right) %
\right] ,
\end{equation}

given by,%
\begin{eqnarray}
I_{\alpha },\text{\textbf{\ }}^{\nu }\text{ }_{\beta \sigma }\left( \mathbf{%
x,x}{\acute{}}\right) &=&\xi ^{\nu }\left( \mathbf{x}{\acute{}}\right)
\left\{ \left[ A_{\alpha }\left( \mathbf{x}\right) ,\partial {\acute{}}%
_{\beta }A_{\sigma }\left( \mathbf{x}{\acute{}}\right) \right] -\left[
A_{\alpha }\left( \mathbf{x}\right) ,\partial {\acute{}}_{\sigma }A_{\beta
}\left( \mathbf{x}{\acute{}}\right) \right] \right\} +  \notag \\
&&+in_{0\sigma }\xi ^{\nu }\left( \mathbf{x}{\acute{}}\right) \left(
g_{\alpha \beta }-\nabla ^{-2}\partial _{\alpha }\partial {\acute{}}_{\beta
}\right) \delta \left( \mathbf{\vec{x}}-\mathbf{\vec{x}}{\acute{}}\right)
-in_{0\beta }\xi ^{\nu }\left( \mathbf{x}{\acute{}}\right) \left( g_{\alpha
\sigma }-\nabla ^{-2}\partial _{\alpha }\partial {\acute{}}_{\sigma }\right)
\delta \left( \mathbf{\vec{x}}-\mathbf{\vec{x}}{\acute{}}\right) .
\end{eqnarray}

The tensor $\mathcal{T}_{\alpha \beta }^{b}\left( x\right) $ can be written
as,

\begin{eqnarray}
\mathcal{T}_{\alpha \beta }^{b}\left( x\right) &=&\lim_{x\rightarrow x{%
\acute{}}}\left[ \left( U_{\alpha \beta }^{b}\left( x\right) -\mathcal{V}%
_{\alpha \beta }^{b}\left( x\right) \right) \left( g_{\lambda \psi }+\frac{%
\partial _{\lambda }\partial {\acute{}}_{\psi }}{\Box }\right) \frac{1}{\Box 
}\right] ,  \notag \\
U_{\alpha \beta }^{b}\left( x\right) &=&\delta _{\nu }^{\lambda }\Sigma
^{\sigma }\text{ }_{\alpha }\delta _{\sigma }^{\psi }\Sigma {\acute{}}^{\nu }%
\text{ }_{\beta }-\delta _{\nu }^{\lambda }\Sigma ^{\sigma }\text{ }_{\alpha
}\delta _{\beta }^{\psi }\Sigma {\acute{}}^{\nu }\text{ }_{\sigma }-\delta
_{\alpha }^{\lambda }\Sigma ^{\sigma }\text{ }_{\nu }\delta _{\sigma }^{\psi
}\Sigma {\acute{}}^{\nu }\text{ }_{\beta }+\delta _{\alpha }^{\lambda
}\Sigma ^{\sigma }\text{ }_{\nu }\delta _{\beta }^{\psi }\Sigma {\acute{}}%
^{\nu }\text{ }_{\sigma },  \notag \\
\mathcal{V}_{\alpha \beta }^{b}\left( x\right) &=&\frac{1}{2}g_{\alpha \beta
}\left( \delta _{\nu }^{\lambda }\Sigma ^{\sigma }\text{ }_{\rho }\delta
_{\sigma }^{\psi }\Sigma {\acute{}}^{\nu \rho }-\delta _{\nu }^{\lambda
}\Sigma ^{\sigma }\text{ }_{\rho }\delta ^{\psi \rho }\Sigma {\acute{}}^{\nu
}\text{ }_{\sigma }-\delta _{\nu }^{\lambda }\Sigma ^{\sigma }\text{ }_{\nu
}\delta _{\sigma }^{\psi }\Sigma {\acute{}}^{\nu \rho }+\delta _{\rho
}^{\lambda }\Sigma ^{\sigma }\text{ }_{\nu }\delta ^{\psi \rho }\Sigma {%
\acute{}}^{\nu }\text{ }_{\sigma }\right) ,
\end{eqnarray}

that considering only the first term of the propagator, after some
operations\ $\mathcal{T}_{\alpha \beta }=\mathcal{T}_{\alpha \beta }^{a}+%
\mathcal{T}_{\alpha \beta }^{b},$where,

\begin{eqnarray}
\text{\ } &<&\mathcal{T}_{\alpha \beta }(x)>=\lim_{x\rightarrow x{\acute{}}}%
\left[ \Gamma _{\alpha \beta }\left( \mathbf{x,x}{\acute{}}\right)
G_{0}\left( \mathbf{x,x}{\acute{}}\right) +2i\left( n_{0\alpha }n_{0\beta }-%
\frac{1}{2}g_{\alpha \beta }\delta \left( \mathbf{\vec{x}}-\mathbf{\vec{x}}{%
\acute{}}\right) \right) \right] ;\text{ } \\
\mathcal{iD}_{\alpha \beta } &=&<0\left\vert T\left[ A_{\alpha }\left( 
\mathbf{x}\right) A_{\beta }\left( \mathbf{x}{\acute{}}\right) \right]
\right\vert 0>=g_{\alpha \beta }G_{0}\left( \mathbf{x,x}{\acute{}}\right)
\end{eqnarray}

with

\begin{eqnarray}
\text{\ }G_{0}\left( \mathbf{x,x}{\acute{}}\right) &=&\frac{1}{4\pi ^{2}i}%
\frac{1}{\left( \mathbf{x-x}{\acute{}}\right) }, \\
\mathcal{iD}_{\alpha \beta } &=&<0\left\vert T\left[ A_{\alpha }\left( 
\mathbf{x}\right) A_{\beta }\left( \mathbf{x}{\acute{}}\right) \right]
\right\vert 0>=g_{\alpha \beta }G_{0}\left( \mathbf{x,x}{\acute{}}\right) .
\end{eqnarray}

For dependent fields in a parameter $\epsilon $, the renormalized
energy-momentum tensor is:

\begin{equation}
\text{\ }<\mathcal{T}_{\alpha \beta }(x;\epsilon )>=-i\lim_{x\rightarrow x{%
\acute{}}}\left[ \Gamma _{\alpha \beta }\left( \mathbf{x,x}{\acute{}}\right)
G_{0}\left( \mathbf{x,x}{\acute{}};\epsilon \right) +2i\left( n_{0\alpha
}n_{0\beta }-\frac{1}{2}g_{\alpha \beta }\delta \left( \mathbf{\vec{x}}-%
\mathbf{\vec{x}}{\acute{}}\right) \right) \right] ;\text{ }
\end{equation}

or in another way,

\begin{eqnarray}
\text{\ } &<&\mathcal{T}_{\alpha \beta }(x;\epsilon )>=<\mathcal{T}%
_{\alpha\beta }(x;\epsilon )>-<\mathcal{T}_{\alpha \beta }(x)>; \\
&<&\mathcal{T}_{\alpha \beta }(x;\epsilon )>=-i\lim_{x\rightarrow x{\acute{}}%
}\left[ \Gamma _{\alpha \beta }\left( \mathbf{x,x}{\acute{}}\right)
G_{0}\left( \mathbf{x,x}{\acute{}};\epsilon \right) \right]
\end{eqnarray}

which can be written in terms of $\Gamma $

\begin{equation}
\Gamma _{\alpha \beta }\left( \mathbf{x,x}{\acute{}}\right) =2\xi _{\nu }\xi 
{\acute{}}^{\nu }\left( \partial _{\alpha }\partial {\acute{}}_{\beta }-%
\frac{1}{2}g_{\alpha \beta }\partial _{\alpha }\partial {\acute{}}^{\beta
}\right) -2\left( \xi ^{\sigma }\partial {\acute{}}_{\sigma }\partial
_{\alpha }\xi {\acute{}}_{\beta }\right) ,
\end{equation}

\section{ The eletromagnetism with Lorentz Breaking Even of SME Sector $%
\protect\alpha $-energy-momentum tensor}

In this section we calculate energy-momentum for our model in a compactified
in a toroidal topology. We define the physical (renormalized)
energy-momentum tensor by%
\begin{equation}
\mathcal{T}^{\mu \nu }(x;\alpha )=\left\langle T^{\mu \nu }(x;\alpha
)\right\rangle _{0}-\langle T^{\mu \nu }(x)\rangle _{0}  \label{june123}
\end{equation}%
where $\left\langle T^{\mu \nu }(x;\alpha )\right\rangle _{0}=\langle
0|T^{\mu \nu }(x,\alpha )|0\rangle \equiv \langle \alpha |T^{\mu \nu
}(x)|\alpha \rangle $. This leads to 
\begin{equation*}
\mathcal{T}^{\mu \nu }(x;\alpha )=-i\lim_{x\rightarrow x{\acute{}}}\Gamma
^{\mu \nu }(x,x^{\prime })G_{0}(x-x^{\prime };\alpha ),
\end{equation*}%
where%
\begin{align*}
\overline{G}(x-x^{\prime };\alpha )& =G_{0}(x-x^{\prime };\alpha
)-G_{0}(x-x^{\prime }) \\
& =\int \frac{d^{4}k}{(2\pi )^{4}}e^{-ik(x-x^{\prime })}v^{2}(k_{\alpha
},\alpha )[G_{0}(k)-G_{0}^{\ast }(k)].
\end{align*}

Let us calculate, as an example, the case of temperature defined by $\alpha
=(\beta ,0,0,0)$, with $v^{2}(k_{0};\beta )$ given by Eq.~(\ref{june121}).
Then we have

\begin{equation}
\mathcal{T}_{\alpha \beta }(\beta )=-\frac{\kappa \pi ^{2}}{90n_{0}^{4}}%
\left\{ \xi ^{2}\left( \frac{8n_{0\alpha }n_{0\beta }}{n_{0}^{2}}+5g_{\alpha
\beta }\right) \right\} ,
\end{equation}

taking into account the $\mathcal{T}_{\alpha \beta }^{a}$ term and obtaining
its contribution we have the complete expression of the energy-momentum
tensor\emph{,}

\begin{eqnarray}
\mathcal{T}_{\alpha \beta }(\beta ) &=&-\frac{\pi ^{2}}{45\beta ^{4}}\left(
1-\frac{\kappa }{2}\xi ^{2}\right) \left( g_{\alpha \beta }-4n_{0\alpha
}n_{0\beta }\right) +  \notag \\
&&-\frac{\kappa \pi ^{2}}{90n_{0}^{4}\beta ^{4}}\left\{ \xi ^{2}\left( \frac{%
8n_{0\alpha }n_{0\beta }}{n_{0}^{2}}+5g_{\alpha \beta }\right) +\frac{8}{%
n_{0}^{2}}\left( \xi ^{\sigma }n_{0\sigma }n_{0\alpha }\xi _{\beta }+\xi
^{\sigma }n_{0\sigma }\xi _{\alpha }n_{0\beta }\right) -4\xi _{\alpha }\xi
_{\beta }\right\} ,
\end{eqnarray}

\bigskip where $n_{0}^{\mu }=(1,0,0,0)$. Using the Riemann Zeta function 
\begin{equation*}
\zeta (4)=\sum_{l=1}^{\infty }\frac{1}{l^{4}}=\frac{\pi ^{4}}{90},
\end{equation*}%
we obtain

\begin{equation}
\mathcal{T}_{00}(\beta )=\frac{3\pi ^{2}}{45\beta ^{4}}\left( 1-\frac{\kappa 
}{2}\xi ^{2}\right) -\frac{\kappa \pi ^{2}}{90\beta ^{4}}\left\{ 13\xi ^{2}+%
\frac{16\xi ^{0}\xi _{0}}{n_{0}^{2}}-4\xi _{0}\xi _{0}\right\} ,
\end{equation}

This leads to the Stephan-Boltzmann law for our model, since the energy and
pressure are given respectively by,%
\begin{equation}
E(\beta )=\mathcal{T}_{00}(\beta )=\left\{ 
\begin{array}{ccc}
\left( 1-\frac{\kappa }{2}\right) \frac{\pi ^{2}}{15\beta ^{4}}-25\frac{%
\kappa \pi ^{2}}{90\beta ^{4}}, & \mathrm{for} & \xi _{\rho }\text{ time-like%
} \\ 
&  &  \\ 
\left( 1+\frac{\kappa }{2}\right) \frac{\pi ^{2}}{15\beta ^{4}}-13\frac{%
\kappa \pi ^{2}}{90\beta ^{4}}, & \mathrm{for} & \xi _{\rho }\text{
space-like}%
\end{array}%
\right.  \label{3.19d}
\end{equation}

and%
\begin{equation}
P(\beta )=\mathcal{T}^{33}(\beta )=\left\{ 
\begin{array}{ccc}
\left( 1-\frac{\kappa }{2}\right) \frac{\pi ^{2}}{45\beta ^{4}}+\frac{\kappa
\pi ^{2}}{10\beta ^{4}}, & \mathrm{for} & \xi _{\rho }\text{ time-like} \\ 
&  &  \\ 
\left( 1+\frac{K}{2}\right) \frac{\pi ^{2}}{45\beta ^{4}}-\frac{\kappa \pi
^{2}}{90\beta ^{4}}, & \mathrm{for} & \xi _{\rho }\text{ space-like}%
\end{array}%
\right.
\end{equation}

In the next section, we use a similar procedure to calculate the Casimir
effect.

\section{The Casimir effect for the model}

Initially we consider the Casimir effect at zero temperature. This is given
by our energy-momentum tensor $\mathcal{T}^{\mu \nu }(x;\alpha )$ given in
Eq.~(\ref{june123}), where $\alpha $ accounts for spatial compactifications.
We take $\alpha =(0,0,0,iL)$, with $L$ being the circumference of $S^{1}$.
The Bogoliubov transformation is given in Eq.~(\ref{june125}), that in the
present notation reads 
\begin{equation*}
v^{2}(k_{3};L)=\sum_{l_{3}=1}^{\infty }e^{-iLk^{3}l_{3}}.
\end{equation*}%
Thus $\mathcal{T}^{\mu \nu }(x;L)$ is given by%
\begin{eqnarray}
\mathcal{T}_{\alpha \beta }(L) &=&-\frac{\pi ^{2}}{45\left( n_{3}L\right)
^{4}}\left( 1-\frac{\kappa }{2}\xi ^{2}\right) \left( g_{\alpha \beta
}+4n_{3\alpha }n_{3\beta }\right) -\frac{\kappa \pi ^{2}}{90\left(
n_{3}L\right) ^{4}}X_{\alpha \beta }(L)  \notag \\
X_{\alpha \beta }(L) &=&\xi ^{2}(\frac{8n_{3\alpha }n_{3\beta }}{\left(
n_{3}\right) ^{2}}+5g_{\alpha \beta })+\frac{8}{\left( n_{3}\right) ^{2}}%
\left( \xi ^{\sigma }n_{3\sigma }n_{3\alpha }\xi _{\beta }+\xi ^{\sigma
}n_{3\sigma }\xi _{\alpha }n_{3\beta }\right) -4\xi _{\alpha }\xi _{\beta },
\end{eqnarray}%
where $n_{3\alpha }=(0,0,0,1)$.

For the electromagnetic field with Lorentz breaking symmmetry, the Casimir
effect is calculated for plates apart from each other by a distance $a$,
that is related to \thinspace $L$\cite{kha1}. We consider this fact for the
sake of comparasion. The Casimir energy

\begin{equation}
E(\beta )=\mathcal{T}_{00}(\beta )=\left\{ 
\begin{array}{ccc}
-\left( 1-\frac{\kappa }{2}\right) \frac{\pi ^{2}}{45}\frac{1}{L^{4}}-\frac{%
\kappa \pi ^{2}}{90}\frac{1}{L^{4}}, & \mathrm{for} & \xi _{\rho }\text{
time-like} \\ 
&  &  \\ 
-\left( 1+\frac{\kappa }{2}\right) \frac{\pi ^{2}}{45}\frac{1}{L^{4}}+\frac{%
\kappa \pi ^{2}}{10}\frac{1}{L^{4}}, & \mathrm{for} & \xi _{\rho }\text{
space-like}%
\end{array}%
\right.
\end{equation}

and pressure, respectively, are then given by%
\begin{equation}
P(\beta )=\mathcal{T}^{33}(\beta )=\left\{ 
\begin{array}{ccc}
-\left( 1-\frac{\kappa }{2}\right) \frac{\pi ^{2}}{15}\frac{1}{L^{4}}-\frac{%
17\kappa \pi ^{2}}{90L^{4}} & \mathrm{for} & \xi _{\rho }\text{ time-like}
\\ 
&  &  \\ 
-\left( 1+\frac{\kappa }{2}\right) \frac{\pi ^{2}}{15}\frac{1}{L^{4}}-\frac{%
23\kappa \pi ^{2}}{90}\frac{1}{L^{4}}, & \mathrm{for} & \xi _{\rho }\text{
space-like}%
\end{array}%
\right.
\end{equation}

It is interesting to compare such a result with the Casimir effect for the
electromagnetic field. For the electromagnetic field with Lorentz breaking
symmetry, the Casimir energy, at $T=0$ $K$ and for $\xi _{\rho }$ time-lihe,
it is exactly the same for the electromegnetic field without the extended
therm, eqs 14-16 , $E(T)=-\pi ^{2}/720a^{4}$, where $L=2a$, while for $\xi
_{\rho }$ space-like, this result is to increase of , $E(T)=-4\kappa \pi
^{2}/720a^{4}$. Already for Casimir pressure, as for $\xi _{\rho }$
time-like as well for space-like, the extended therm contributes to the
final result.

The effect of temperature is introduced by taken $\alpha =(i\beta ,0,0,iL)$.
Using Eq. (\ref{june124}), $v^{2}(k^{0},k^{3};\beta ,L)\ $\ is given by%
\begin{align}
v^{2}(k^{0},k^{3};\beta ,L)& =v^{2}(k^{0};\beta
)+v^{2}(k^{3};L)+2v^{2}(k^{0};\beta )v^{2}(k^{3};L)  \notag \\
& =\sum\limits_{l_{0}=1}^{\infty }e^{-\beta
k^{0}l_{0}}+\sum_{l_{3}=1}^{\infty
}e^{-iLk^{3}l_{3}}+2\sum\limits_{l_{0},l_{3}=1}^{\infty }e^{-\beta
k^{0}l_{0}-iLk^{3}l_{3}}.
\end{align}%
The two parts of the energy-momentum tensor is $\mathcal{T}_{\alpha \beta
}^{A}(\beta ,L)$ and $\mathcal{T}_{\alpha \beta }^{B}(\beta ,L)$, 
\begin{eqnarray}
\mathcal{T}_{\alpha \beta }^{A}(\beta ,L) &=&-\frac{2}{\pi ^{2}}\left( 1-%
\frac{\kappa }{2}\xi ^{2}\right) \left\{ \sum_{l_{0}=1}^{\infty }\mathcal{X}%
_{\alpha \beta }^{A}(\beta ,L)+\sum_{l_{3}=1}^{\infty }\mathcal{Y}_{\alpha
\beta }^{A}(\beta ,L)+4\sum_{l_{0},l_{3}=1}^{\infty }\mathcal{Z}_{\alpha
\beta }^{A}(\beta ,L)\right\} ,  \notag \\
\mathcal{X}_{\alpha \beta }^{A}(\beta ,L) &=&\frac{g_{\alpha \beta
}-4n_{0\alpha }n_{0\beta }}{\left( \beta l_{0}\right) ^{4}},\text{ }\mathcal{%
Y}_{\alpha \beta }^{A}(\beta ,L)=\frac{g_{\alpha \beta }-4n_{0\alpha
}n_{0\beta }}{\left( Ll_{3}\right) ^{4}},  \notag \\
\mathcal{Z}_{\alpha \beta }^{A}(\beta ,L) &=&\frac{\left( \beta l_{0}\right)
^{2}\left[ g_{\alpha \beta }-4n_{0\alpha }n_{0\beta }\right] +\left(
2Ll_{3}\right) ^{2}\left[ g_{\alpha \beta }+4n_{3\alpha }n_{3\beta }\right] 
}{\left[ \left( \beta l_{0}\right) ^{2}+\left( Ll_{3}\right) ^{2}\right] ^{3}%
},\text{ }
\end{eqnarray}

and%
\begin{eqnarray}
\mathcal{T}_{\alpha \beta }^{B}(\beta ,L) &=&\frac{-2K}{\pi ^{2}}%
\sum_{l_{0},l_{3}=1}^{\infty }\xi ^{2}\left( \frac{-8\left[ \left(
Ll_{3}\right) ^{2}n_{3\alpha }n_{3\beta }-\left( \beta l_{0}\right)
^{2}n_{0\alpha }n_{0\beta }\right] }{\left[ \left( Ll_{3}\right) ^{2}-\left(
\beta l_{0}\right) ^{2}\right] ^{3}}+\frac{5g_{\alpha \beta }\left[ \left(
Ll_{3}\right) ^{2}-\left( \beta l_{0}\right) ^{2}\right] }{\left[ \left(
Ll_{3}\right) ^{2}-\left( \beta l_{0}\right) ^{2}\right] ^{3}}\right) + 
\notag \\
&&+16\frac{\left[ \left( Ll_{3}\right) ^{2}\xi ^{\sigma }n_{3\sigma
}n_{3\alpha }\xi _{\beta }-\left( \beta l_{0}\right) ^{2}\xi ^{\sigma
}n_{0\sigma }\xi _{\beta }n_{0\alpha }\right] }{\left[ \left( Ll_{3}\right)
^{2}-\left( \beta l_{0}\right) ^{2}\right] ^{3}}-\frac{4\xi _{\alpha }\xi
_{\beta }}{\left[ \left( Ll_{3}\right) ^{2}-\left( \beta l_{0}\right) ^{2}%
\right] }
\end{eqnarray}

Then taking into account the total expression of the energy-momentu tensor
is $\mathcal{T}_{\alpha \beta }(\beta ,L)=\mathcal{T}_{\alpha \beta
}^{A}(\beta ,L)+\mathcal{T}_{\alpha \beta }^{B}(\beta ,L)$, we evaluete the
Casimir energy $E(\beta ,L)$ and the Casimir pressure $P(\beta ,L)$ with the
temperature dependence. The Casimir energy $\mathcal{T}^{00}(\beta )$ and
pressure $\mathcal{T}^{33}(\beta )$ are given in the case of $\xi ^{\mu
}=(1;0,0,0)$ , respectively by 
\begin{eqnarray}
E(\beta ,L) &=&\left( 1-\frac{\kappa }{2}\right) \left( A(\beta ,L)+\frac{8}{%
\pi ^{2}}\sum_{l_{0},l_{3}=1}^{\infty }B(\beta ,L)\right) +\left( 1-\frac{%
\kappa }{2}\right) \left( \frac{2K}{\pi ^{2}}\sum_{l_{0},l_{3}=1}^{\infty
}C(\beta ,L)\right) ,  \notag \\
A(\beta ,L) &=&\frac{\pi ^{2}}{45}\left( \frac{3}{\beta ^{4}}-\frac{1}{L^{4}}%
\right) ,  \notag \\
B(\beta ,L) &=&\frac{3\left( \beta l_{0}\right) ^{2}-4\left( Ll_{3}\right)
^{2}}{\left[ \left( \beta l_{0}\right) ^{2}+\left( Ll_{3}\right) ^{2}\right]
^{3}},  \notag \\
C(\beta ,L) &=&\frac{8\left( \beta l_{0}\right) ^{2}}{\left[ \left(
Ll_{3}\right) ^{2}-\left( \beta l_{0}\right) ^{2}\right] ^{3}}-\frac{5}{%
\left[ \left( Ll_{3}\right) ^{2}-\left( \beta l_{0}\right) ^{2}\right] ^{2}}+%
\frac{4}{\left[ \left( Ll_{3}\right) ^{2}-\left( \beta l_{0}\right) ^{2}%
\right] },
\end{eqnarray}

and 
\begin{eqnarray}
P(\beta ,L) &=&\left( 1-\frac{\kappa }{2}\right) \left( D(\beta ,L)+\frac{8}{%
\pi ^{2}}\sum_{l_{0},l_{3}=1}^{\infty }F(\beta ,L)+\frac{2K}{\pi ^{2}}%
\sum_{l_{0},l_{3}=1}^{\infty }G(\beta ,L)\right) ,  \notag \\
D(\beta ,L) &=&\frac{\pi ^{2}}{45}\left( \frac{1}{\beta ^{4}}-\frac{3}{L^{4}}%
\right) ,  \notag \\
F(\beta ,L) &=&\frac{3\left( \beta l_{0}\right) ^{2}-12\left( Ll_{3}\right)
^{2}}{\left[ \left( \beta l_{0}\right) ^{2}+\left( Ll_{3}\right) ^{2}\right]
^{3}},  \notag \\
G(\beta ,L) &=&\frac{8\left( Ll_{3}\right) ^{2}}{\left[ \left( Ll_{3}\right)
^{2}-\left( \beta l_{0}\right) ^{2}\right] ^{3}}+\frac{5}{\left[ \left(
Ll_{3}\right) ^{2}-\left( \beta l_{0}\right) ^{2}\right] ^{2}}.
\end{eqnarray}

The Casimir energy and pressure are given in the case of $\xi ^{\mu
}=(0;0,0,1)$ , respectively by 
\begin{eqnarray}
E(\beta ,L) &=&\left( 1+\frac{\kappa }{2}\right) \left( H(\beta ,L)+\frac{8}{%
\pi ^{2}}\sum_{l_{0},l_{3}=1}^{\infty }I(\beta ,L)+\frac{2K}{\pi ^{2}}%
\sum_{l_{0},l_{3}=1}^{\infty }J(\beta ,L)\right) ,  \notag \\
H(\beta ,L) &=&\frac{\pi ^{2}}{45}\left( \frac{3}{\beta ^{4}}-\frac{1}{L^{4}}%
\right) ,  \notag \\
I(\beta ,L) &=&\frac{3\left( \beta l_{0}\right) ^{2}-4\left( Ll_{3}\right)
^{2}}{\left[ \left( \beta l_{0}\right) ^{2}+\left( Ll_{3}\right) ^{2}\right]
^{3}},  \notag \\
J(\beta ,L) &=&\frac{-8\left( \beta l_{0}\right) ^{2}}{\left[ \left(
Ll_{3}\right) ^{2}-\left( \beta l_{0}\right) ^{2}\right] ^{3}}+\frac{5}{%
\left[ \left( Ll_{3}\right) ^{2}-\left( \beta l_{0}\right) ^{2}\right] ^{2}},
\end{eqnarray}

and 
\begin{eqnarray}
P(\beta ,L) &=&\left( 1+\frac{\kappa }{2}\right) \left( L(\beta ,L)+\frac{8}{%
\pi ^{2}}\sum_{l_{0},l_{3}=1}^{\infty }M(\beta ,L)+\frac{2K}{\pi ^{2}}%
\sum_{l_{0},l_{3}=1}^{\infty }N(\beta ,L)\right) ,  \notag \\
L(\beta ,L) &=&\frac{\pi ^{2}}{45}\left( \frac{1}{\beta ^{4}}-\frac{3}{L^{4}}%
\right) ,  \notag \\
M(\beta ,L) &=&\frac{3\left( \beta l_{0}\right) ^{2}-12\left( Ll_{3}\right)
^{2}}{\left[ \left( \beta l_{0}\right) ^{2}+\left( Ll_{3}\right) ^{2}\right]
^{3}},  \notag \\
N(\beta ,L) &=&\frac{8\left( Ll_{3}\right) ^{2}}{\left[ \left( Ll_{3}\right)
^{2}-\left( \beta l_{0}\right) ^{2}\right] ^{3}}-\frac{5}{\left[ \left(
Ll_{3}\right) ^{2}-\left( \beta l_{0}\right) ^{2}\right] ^{2}}-\frac{4}{%
\left[ \left( Ll_{3}\right) ^{2}-\left( \beta l_{0}\right) ^{2}\right] }.
\end{eqnarray}%
The first two terms of these expressions are, respectively, the
Stephan-Boltzmann term and the Casimir effect at $T=0$. The last term
accounts for the simultaneous effect of spatial compactification, described
by $L$, and temperature, $T=1/\beta $.

\section{Concluding remarks}

In this work, we have adopted the Thermo-Field Dynamics (TFD) approach, a
real-time formalism for Quantum Field Theory at finite temperatures, to
study the Casimir Effect in the framework of an electrodynamical model with
the so-called even Lorentz-symmetry breaking term in the photon sector of
the SME. We have initially worked out the expression for the energy-momentum
tensor of the model in terms of the TFD propagator, by considering an $%
S^{1}\times S^{1}\times R^{2}$-topology, where the two factors $S^{1}$
correspond to a compactified space-like coordinate and the finite
temperature. The TFD techanalities are very appropriated to deal with the
renormalized energy-momentum tensor of the model with LSV under
consideration. The Casimir energy and pressure have been both explicitly
calculated and we find that the corrections that arise directly from the LSV
come out linear in the $\kappa $-parameter, which is constrained by several
tests. The expressions we calculate for the energy and pressure can be used
for the attainment of a new category of bound on the $\kappa $-parameter.
Different cases have been considered where the external vector responsible
for the LSV may be time- and space-like. Though not explicitly mentioned,
the light-like case also gives rise to non-trivial effects on the energy and
pressure. In the case in which the external vector responsible for the LSV
is time-like the Casimir effect do not present influence of the Lorentz
breaking.

Clearly, the tests require extremely high-precision measuremts, once the
known limits on $\kappa $ are very tiny. However, for very high
temperatures, the combined effect between the LSV parameter and the
temperature itself may yield a measurable effect on the energy and pressure.
On the other hand, one might adopt current measurements of the energy and
pressure at finite temperatures to set up a new class of limits on the $%
\kappa $-parameter. A point which remains to be investigated in the LSV
scenario we are considering is the thermal Casimir effect in the interaction
of graphene with a metal. In this type of system we could enhance the effect
of LSV and we could end up with more stringent constraints on the LSV
parameters.

\textbf{Ackowledgements}

The authors are indebted and express their gratitude to the colleague Ademir
Santana for participation and decisive discussions at an early stage of the
present paper. The Authors also thank CNPq and CAPES (of Brazil) for
financial support.

\end{document}